\documentclass[aps,prd,twocolumn]{revtex4-1}

\usepackage{amsmath,amssymb,amsfonts}
\usepackage{dcolumn,color,latexsym,placeins}

\newcommand{\be}{\begin{eqnarray}}
\newcommand{\ee}{\end{eqnarray}}
\newcommand{\ba}{\begin{eqnarray}}
\newcommand{\ea}{\end{eqnarray}}

\def\ba{\begin{eqnarray}}
\def\ea{\end{eqnarray}}
\def\be{\begin{equation}}
\def\ee{\end{equation}}

\def\nn{\nonumber}

\def\eps{\epsilon}

\newcommand{\calR}{{\cal R}}
\newcommand{\calH}{{\cal H}}
\newcommand{\lapl}{\mathop\Delta^{(3)}}
\def\R{{\mathcal R}}

\begin{document}

\begin{flushright}
YITP-16-72
\end{flushright}

\title{Global adiabaticity and non-Gaussianity consistency condition}

\author{Antonio Enea Romano,$^{1,3}$ Sander Mooij$^2$ and Misao Sasaki$^3$}
\affiliation
{
${}^{1}$Instituto de Fisica, Universidad de Antioquia, A.A.1226, Medellin, Colombia\\
%${}^{1}$Department of Physics, University of Crete, 71003 Heraklion,Greece \\
${}^{2}$Grupo de Cosmolog\'ia y Astrof\'isica Te\'orica,
Departamento de F\'{i}sica, FCFM, Universidad de Chile,
Blanco Encalada 2008, Santiago, Chile\\
${}^{3}$Center for Gravitational Physics,
Yukawa Institute for Theoretical Physics, Kyoto University, Kyoto 606-8502, Japan\\
%${}^{3}$Department of Physics, University of Torino, Via P. Giuria 1, I--10125 Torino, Italy\\
%${}^{4}$INFN, Sezione di Torino, Via P. Giuria 1, I--10125 Torino, Italy
}

\begin{abstract}
In the context of single-field inflation, 
the conservation of the curvature perturbation on comoving slices, $\R_c$,
on super-horizon scales is one of the assumptions necessary to derive the 
consistency condition between the squeezed limit of the bispectrum and
 the spectrum of the primordial curvature perturbation. 
However, the conservation of $\R_c$ holds only after the perturbation has reached
the adiabatic limit where the constant mode of $\R_c$ dominates over the other 
(usually decaying) mode. In this case, the non-adiabatic pressure perturbation
defined in the thermodynamic sense,
 $\delta P_{nad}\equiv\delta P-c_w^2\delta\rho$ where $c_w^2=\dot P/\dot\rho$,
usually becomes also negligible on superhorizon scales. Therefore one might
think that the adiabatic limit is the same as thermodynamic adiabaticity.
This is in fact not true. In other words, thermodynamic adiabaticity is
not a sufficient condition for the conservation of $\R_c$ on super-horizon scales.
In this paper, we consider models that satisfy $\delta P_{nad}=0$ on all scales,
which we call global adiabaticity (GA), which is guaranteed if $c_w^2=c_s^2$, 
where $c_s$ is the phase velocity of the propagation of the perturbation.
A known example is the case of ultra-slow-roll(USR) inflation in which $c_w^2=c_s^2=1$. In order to generalize USR we develop 
a method to find the Lagrangian of GA K-inflation models from the behavior of background quantities as functions of the scale factor. 
Applying this method we show that there indeed exists a wide 
class of GA models with $c_w^2=c_s^2$, which allows $\R_c$ to grow on 
superhorizon scales, and hence violates the non-Gaussianity consistency condition.
\end{abstract}

\maketitle

\section{Introduction}
A period of accelerated expansion during the early stages of the evolution
 of the Universe, called inflation \cite{infl,infl2,infl3}, is able to account for several otherwise 
difficult to explain features of the observed Universe such the high level
 of isotropy of the cosmic microwave background (CMB) \cite{Ade:2015lrj} radiation
 and the small value of the curvature.
Some of the simplest inflationary  models are  based on a single slowly-rolling scalar 
field, and they are in good agreement with observations. It is commonly assumed 
in slow-roll models that adiabaticity in the thermodynamic sense, 
$\delta P_{nad}\equiv \delta P-c_w^2\delta\rho=0$ where $c_w^2=\dot P/\dot\rho$,
implies the conservation of the curvature perturbation on uniform density 
slices $\zeta$, and hence the conservation of the curvature perturbation on comoving slices $\R_c$, on super-horizon scales.

In \cite{Romano:2015vxz} it was shown that there  can be  important exceptions, i.e.
 in some cases thermodynamic adiabaticity does not necessarily imply the 
super-horizon conservation of $\R_c$ and $\zeta$, and that they can differ from 
each other. 
This can happen even for models in which $c_w^2=c_s^2$.
An example is  ultra-slow-roll (USR) 
inflation \cite{Tsamis:2003px,Kinney:2005vj},
 which has exact adiabaticity $\delta P_{nad}=0$ on all scales.
In USR inflation, both $\R_c$ and $\zeta$ 
exhibit super-horizon growth but their behavior is very different from each other. As has been stressed in \cite{Namjoo:2012aa}, the non-freezing of $\R_c$ has important phenomenological consequences. Since the freezing of $\R_c$ on superhorizon scales is a necessary ingredient \cite{CreZar} for Maldacena's consistency relation \cite{Maldacena} to hold, models that do not conserve $\R_c$ can actually violate that consistency condition.

In this paper focusing on K-inflation, i.e., Einstein-scalar models with 
a general kinetic term, we explore in a general way other single field models 
which have $c_w^2=c_s^2$, hence satisfy $\delta P_{nad}=0$ on all scales
which we call globally adiabatic (GA), but which may not conserve $\R_c$. 
We find a generalization of the USR model. A different generalization 
without imposing the condition $c_w^2=c_s^2$ was discussed in \cite{Chen:2013aj,Martin:2012pe}.

The method we adopt is based on establishing a general condition for the 
non-conservation of $\R_c$ in terms of the dependence of the background 
quantities, in particular the slow-roll parameter $\epsilon\equiv -\dot H/H^2$ 
and the sound velocity $c_s$, on the scale factor $a$.

We first derive the necessary condition for the co\-mo\-ving curvature perturbation 
$\R_c$ to grow on superhorizon scales. Next we determine $\rho(a)$ and $P(a)$
 by solving the continuity equation.
Then using the equivalence between barotropic fluids and $K$-inflationary 
models which satisfy the condition $c_w^2=c_s^2$ \cite{Arroja:2010wy,Unnikrishnan:2010ag},
 we determine the corresponding Lagrangian for the equivalent 
scalar field model. Using this method we obtain a new class of GA scalar field 
models which do not conserve $\R_c$. 

Throughout the paper we denote the proper-time derivative by a dot
($\dot{~}=d/dt$), the conformal-time derivative by a prime 
($\prime{~}=d/d\eta=a\,d/dt$) and the Hubble expansion rates in proper
and conformal times by $H=\dot a/a$ and $\calH=a'/a$, respectively.
We also use the terminology ``adiabaticity'' for thermodynamic adiabaticity $\delta P_{nad}=0$ throughout the paper.

\section{Conservation of $\R_c$ and global adiabaticity}
We set the perturbed metric as 
\begin{eqnarray}
ds^2&=&a^2\Bigl[-(1+2A)d\eta^2+2\partial_jB dx^jd\eta\nn\\
&& \qquad
+\left\{\delta_{ij}(1+2\calR)+2\partial_i\partial_j E\}dx^idx^j\right\}
\Bigr]\,.
\label{metric}
\end{eqnarray}

In \cite{Romano:2015vxz} it was shown that  independently of the gra\-vi\-ty
 theory and for  generic matter  the energy-momentum conservation 
equations imply
\ba
\delta P_{nad}=\left[\left(\frac{c_w}{c_s}\right)^2-1\right]
 (\rho+P) A_c \label{dpnadAc} \,,
\ea
where the subscript $c$ means a quantity evaluated on comoving slices
defined by $\delta T^i_0=0$ (or equivalently slices on which the scalar field
is homogeneous). In the case of general relativity, the additional 
relation $A_c=\dot\R_c/H$ gives an important relation for the time 
derivative of $\R_c$
\ba
\delta P_{ nad}=
\left[\left(\frac{c_w}{c_s}\right)^2-1\right](\rho+P)\frac{\dot{\R_c}}{H} 
. \label{dpnad} 
\ea

The non-adiabatic pressure perturbation is given according to its 
thermodynamics definition
\be
\delta P_{nad}\equiv \delta P - c_w^2 \delta\rho \label{dpnadther} .
\ee
This definition of $\delta P_{nad}$ is important because 
it is gauge invariant and $\delta P_{nad}=\delta P_{ud}$, 
where $\delta P_{ud}$ is the pressure perturbation on uniform density slices.
It appears in the equation for the curvature perturbation on uniform density slices 
$\zeta\equiv\R_{ud}$ obtained from the energy conservation law~\cite{Wands:2000dp},
\ba
\zeta'=-\frac{\calH\delta P_{nad}}{(\rho+P)}+\frac{1}{3}\lapl\left(v-E'\right)_{ud}\,
\label{dzeta}
\ea
where $v$ is the 3-velocity potential ($v=\delta\phi/\phi'$ for a scalar field).
In general, the curvature perturbations on uniform density and comoving
slices are related as
\ba
\zeta=\R_c+\frac{\delta P_{ nad}}{3(\rho+P)(c^2_s-c^2_w)} 
\label{zetaR} \,.
\ea
A common interpretation of these equations (see for example \cite{lms,Christopherson:2008ry}) is that when
 $\delta P_{nad}=0$ with $c_w^2 \neq c_s^2$, $\zeta$ and $\R_c$ 
are equal because of eq.~(\ref{zetaR}), and they are both conserved 
on super-horizon scales because of eq.~(\ref{dpnad}).

The equation (\ref{dpnad}) is the key relation to understand how 
$\R_c$ depends on the non-adiabatic pressure $\delta P_{nad}$.
First of all let us note that this equation is valid on any scale. 
The advantage of it with respect to eq.~(\ref{dzeta}) is that it does 
not involve gradient terms, so it allows us to directly relate
$\delta P_{\rm nad}$ to $\dot{\R}_{c}$ if $c_w^2\neq c_s^2$, 
while in eq.~(\ref{dzeta}) $\dot{\zeta}$ depends on spatial gradients, 
which in the case of USR are not negligible on super-horizon 
scales \cite{Romano:2015vxz}. 
This explains while in USR in which $c_w^2=c_s^2=1$,
both $\R_c$ and  $\zeta$ are not conserved despite $\delta P_{nad}=0$.

It should be noted here that for slow-roll attractor models $c_w^2\neq c_s^2$
in general, and $\R_c$ is time-varying on sub-horizon scales.
This implies that the non-adiabatic pressure perturbation $\delta P_{nad}$
on sub-horizon scales is not zero. In other words, the attractor models are 
adiabatic only on super-horizon scales, and we call these models super-horizon 
adiabatic (SHA). 

From eq.~(\ref{dpnad}) we can immediately deduce that in ge\-ne\-ral relativity 
there are two possible scenarios for the non-conservation of $\R_c$,
\begin{align}
\mbox{(1)}\quad &c_s^2=c_w^2\,,\quad \delta P_{nad}=0 
 \,,
\nonumber\\
\mbox{(2)}\quad &c_s^2\neq c_w^2\,,\quad\delta P_{nad}\neq 0 \,  \label{GA}.
\end{align}
The second case was studied in \cite{Chen:2013aj, Martin:2012pe}. Here we focus
on the first case. It is trivial to see that because of the gauge 
invariance of $\delta P_{nad}$ the condition $c_w^2=c_s^2$ automatically
implies $\delta P_{nad}=0$. The models satisfying the condition 
$c_s^2-c_w^2=\delta P_{nad}=0$ are adiabatic on any scale, and
 because of this we call them globally adiabatic (GA).
In GA models an explicit calculation can reveal the super-horizon 
behavior of $\R_c$, and $\zeta$, as was shown in \cite{Romano:2015vxz} in the case of USR. 
Below, we develop an inversion method to find a new class of models 
that violate the conservation of $\R_c$ 
without solving the perturbations equations.

\section{Globally adiabatic K-essence models}
 The condition $c_w^2=c_s^2$ has been studied in the context of
 K-inflation \cite{Arroja:2010wy} described by the action
\begin{equation}
S=\frac{1}{2}\int d^4x\sqrt{-g}\left[M^2_{Pl}R+2P(X,\phi)\right] \,,
\label{action}
\end{equation}
and  it was shown that it is satisfied by scalar field models 
with the Lagrangian of the form,
\be
P(X,\phi)=u(X g(\phi))=u(Y)\,. \label{condition}
\ee
These models are
 %indeed globally adiabatic, since they are
  equivalent 
to a barotropic perfect fluid, i.e. a fluid with equation of state $P(\rho)$. See also \cite{Quercellini:2007ht,Faraoni:2012hn,Diez-Tejedor:2013nwa,Wongjun:2016tva}.
We note again that these models \textit{are adiabatic on any scale} (GA), 
contrary to the slow-roll attractor models, which are 
adiabatic \textit{only on super-horizon scales} (SHA). 
The fact that they are mutually exclusive can be readily seen by
considering the hypothetical case of $\delta P_{nad}=0$ {\it and} 
$c_w^2\neq c_s^2$. In this case Eq.~(\ref{dpnad}) which is valid on any scale 
would mean $\R_c$ should be {\it frozen on all scales}. 
In contrast, the condition $c_w^2=c_s^2$ allows for the
curvature perturbation to \textit{evolve} both on \textit{sub-horizon} 
and \textit{super-horizon} scales.

In \cite{Arroja:2010wy} it was shown that is possible to associate 
any barotropic perfect fluid with an equivalent K-inflation model 
according to
\be
2\int^P\frac{du}{F(u)}=\log(Y) \,, \label{fY}
\ee
where $F(P)=\rho(P)+P$ and $Y=g(\phi)X$ with 
$X=-g^{\mu\nu}\partial_\mu\phi\partial_\nu\phi/2$.
These models are the ones which could violate the conservation of $\R_c$
 for adiabatic perturbations, since they satisfy $c_w^2=c_s^2$.  
It is noted of course that the global adiabaticity is not
the sufficient condition for the non-conservation of $\R_c$. Not all 
GA models violate the conservation of $\R_c$ on super-horizon scales.

\section{General conditions for super-horizon growth of $\R_c$}
From the equation for the curvature perturbation on comoving slices,
\be
\frac{\partial}{\partial t}\left(\frac{a^3\eps}{c_s^2}
\frac{\partial}{\partial t}\R_c\right)-a\eps \Delta \R_c=0 \label{eomR2} \,,
\ee
we can deduce, after re-expressing the time derivative in terms 
of the derivative respect to the scale factor $a$, that on superhorizon 
scales there is (apart from a constant solution) a solution of the form,
\ba
\R_c &\propto& \int^a\frac{da}{a} f(a)\,;
\quad 
f(a)\equiv{\frac{c_s^2(a)}{Ha^3 \epsilon(a)}} \,, \label{Rca}
\ea
where we have introduced the function $f(a)$ for later convenience.
In conventional slow-roll inflation $c_s^2$ and $\epsilon$ are
both slowly varying, hence the integral rapidly approaches a constant,
rendering $\R_c$ conserved. The time dependent part of the
above solution corresponds to the decaying mode.

The necessary and sufficient condition for super-horizon freezing is that there exists some $\delta>0$ for which
\be
\lim _{a\to\infty}a^\delta f(a)=0 . \label{freezeF}
\ee
By definition of inflation, $H$ must be sufficiently slowly varying;
$\epsilon=-\dot H/H^2\ll1$. So we may neglect the time dependence of
$H$ in (\ref{Rca}) at leading order, while $\epsilon$ and $c_s^2$ 
may vary rapidly in time. 
For models for which  $\epsilon\approx a^{-n}$ and $c_s^2\approx a^q$ we get 
\be
f \propto a^{q+n-3} \,,
\ee
hence the condition for freezing is 
\be
q+n-3<0 \,.
\ee 
If this condition is violated, i.e. $q+n-3\geq 0$, 
then the solution (\ref{Rca}) will grow on super-horizon scales.
This happens for example in USR., which corresponds to $c_s^2=1$ and
$\epsilon\propto a^{-6}$, i.e. $q=0$, and $n=6$. (The super-horizon growth of $\R_c$ in USR can also be understood as a direct consequence of the non-attractor nature of USR \cite{Mooij:2015yka}.) In general, we expect
that $q$ would not become very large. This implies $\epsilon$ should 
decrease sufficiently rapidly. Conversely, if $\epsilon$ decreases
sufficiently rapidly, then
the growth of $\R_c$ on superhorizon scales will follow.

\section{Barotropic model}

We have shown that GA models could violate the super-horizon 
conservation of $\R_c$, so now we will look for GA K-essence models 
which do indeed violate it, based on the freezing condition in eq.~(\ref{freezeF}).
Inspired by the equivalence between barotropic fluids and GA K-essence 
models \cite{Arroja:2010wy} we will first look for barotropic fluids that can give 
the growing curvature perturbation on superhorizon scales. From the very beginning we will set $c_w^2=c_s^2$.

Using the Friedmann equation we can write the slow-roll 
parameter $\epsilon$ as
\be
\epsilon = -\frac{\dot{H}}{H^2}=\frac{3}{2}\frac{\rho+P}{\rho} \,.
\label{epsilon}
\ee
In terms of the scale factor and $\epsilon$ the energy conservation equation reads
\be
\frac{d\rho}{da}+\frac{3}{a}(\rho+p)
=\frac{d\rho}{da}+\frac{2\epsilon\rho}{a}=0 .
\label{CE}
\ee
We may now define the quantity $b(a)=2\epsilon\rho$. It appears naturally in the continuity equation and plays a crucial role in regards to the super-horizon behavior of curvature perturbations because the function $f(a)$ can be re-written in terms of it as 
\ba
f(a)\propto \frac{H c_s^2}{a^3 b(a)} \,.
\ea
Integrating the energy conservation equation  we get
\ba
\rho(a)=\rho_0\exp\left[-2\int_{a_0}^a{\frac{\epsilon}{a}da}\right]=\int{-\frac{b(a)}{a} da} \,.
\label{rhoa}
\ea
Using eq.~(\ref{epsilon}), we then obtain
\ba
P(a)=\left(\frac{2}{3}\epsilon-1\right)\rho\,. 
\label{Prho} 
\ea

The sound velocity is given by 
\begin{align}
c_w^2=c_s^2=\frac{dP}{d\rho}&=-1+\frac{1}{3}\frac{db(a)}{d\rho}
\nonumber\\
&=-1+\frac{1}{3}\frac{db(a)}{da}\Big/\left(\frac{d\rho}{da}\right)
\nonumber\\
&=-1-\frac{a}{3b(a)}\frac{db(a)}{da}
\,.
\label{cs2}
\end{align}

We now consider the behavior of $f(a)$ introduced in (\ref{Rca}).
As mentioned before, we consider the case when $\epsilon$ decreases
sufficiently rapidly. In this case, $\rho=3H^2M_P^2$ approaches
a constant rapidly. Hence the time dependence of $\rho$ may be neglected
compared to that of other quantities that vary far more rapidly.
With this approximation, assuming $\epsilon\propto a^{-n}$,
we find
\begin{align}
c_s^2\approx\frac{n-3}{3}\,,
\end{align}
which means $q\approx0$, and 
\begin{align}
f(a)={\frac{c_s^2(a)}{Ha^3 \epsilon(a)}}\propto a^{n-3}\,,
\end{align}
which satisfies the condition for the growth if $n>3$, in accordance
with the original anticipation. In passing, it is interesting to
note that the condition $n>3$ implies $c_s^2>0$, a necessary condition
to avoid the gradient instability of the perturbation.
Thus virtually all GA models that are free from the gradient instability
exhibit superhorizon growth of the comoving curvature perturbation $\R_c$.

\section{Scalar field model}

Let us now find a scalar field model that corresponds to the barotropic
model discussed in the previous section.
As a warm-up, let us consider the USR case, whose fluid interpretation has already been studied in \cite{fluid}. In this case, we exactly have
$c_s^2=1$. 
From eq.~(\ref{cs2}), this implies 
$b/2=\epsilon\rho\bigl(=3(\rho+P)/2\bigr)\propto a^{-6}$.
Also $c_s^2=1$ implies $\rho=P+const.$ Inserting this into eq.~(\ref{fY}) gives
\begin{align}
\frac{2dP}{2P+const.}=\frac{dY}{Y}\,.
\end{align}
Thus up to a constant term $P$ and $Y$ are the same,
\begin{align}
P=Y+const..
\end{align}
Absorbing $g(\phi)$ in $Y$ into the definition of the scalar field by
$g^{1/2}d\phi\to d\phi$, this is indeed the Lagrangian for 
a minimally coupled massless scalar with a cosmological constant:
\begin{align}
L=P(\phi,X)=X-V_0\,.   \label{Lag}
\end{align}
This is consistent with $\rho+P=2X\propto\epsilon \rho\propto a^{-6}$.

Let us generalize the USR case.  As in the previous section, we
consider models that have the behavior of $\epsilon\rho$ as
\begin{align}
2\epsilon\rho=b(a)\,,
\label{bdef}
\end{align}
where $b(a)$ should decrease faster than $a^{-3}$
asymptotically at $a\to\infty$ but otherwise is an arbitrary function.
Then we have
\begin{align}
F(P)\equiv \rho+P=2H^2\epsilon=\frac{2\epsilon\rho}{3}=\frac{b(a)}{3}\,,
\label{FPdef}
\end{align}
which gives
\begin{align}
\frac{dY}{Y}=2\frac{dP}{F(P)}=6\frac{dP}{2\epsilon\rho}
=6\frac{dP}{b(a)}\,.
\end{align}

For $dP$, using the energy conservation law,
we may rewrite it as
\begin{align}
dP&=d\left(-\rho +F(P)\right)=-d\rho+\frac{db(a)}{3}
\nonumber\\
&=3\frac{da}{a}(\rho+P)+\frac{db(a)}{3}
=b(a)\frac{da}{a}+\frac{db(a)}{3}\,.
\label{dP}
\end{align}
Therefore we have
\begin{align}
\frac{dY}{Y}=6\frac{dP}{b(a)}=6\frac{da}{a}+2\frac{db}{b}\,.
\label{dlnY}
\end{align}
Hence
\begin{align}
Y\propto a^6b^2\,. \label{Ya}
\end{align}
This is consistent with the USR case in which $b(a) \propto a^{-6}$ 
and $Y=X \propto a^{-6}$.

This relation is quite useful since it allows to rewrite the freezing function $f(a)$ as
\be
f(a)\propto \frac{H c_s^2}{\sqrt{Y}} \,,
\ee
from which we can deduce that $Y(a)$ determines the super-horizon behavior of $\R_c$. In particular, for the mo\-dels we are considering in which  $c_s$ is constant, we infer that super-horizon growth can happen in the limit $Y\rightarrow 0$.

For a given choice of $b(a)$, eq.~(\ref{Ya}) can be inverted to give
the scale factor as a function of $Y$, $a=a(Y)$. 
Also eq.~(\ref{dP}) can be integrated to give $P=P(a)$. Combining these two,
one can obtain the Lagrangian for the scalar field, $L=P=P(Y)$.

Note that in GA models there is a one-to-one correspondence between the scale factor and  state variables such as $P(a)$ and $\rho(a)$, which is the reason why we can also write a barotropic equation of state $P(\rho)=P(a(\rho))$. Once any of the functions $P(a),\rho(a),b(a),\epsilon(a),Y(a)$ is specified, all the others are specified too, as well as  the equation of state $P(\rho)$ or its scalar field equivalent Lagrangian $P(Y)$, which is in fact the basis of the inversion method that we are developing in this paper.

\section{examples}
Here we give a couple of specific K-inflation
models that are globally adiabatic and violate the convervation of $\R_c$. Given the parametric behaviour of $b\equiv 2 \eps\rho$, our inversion method allows us to deduce the Lagrangian.

\subsection{Ex 1: Generalized USR}

Let us consider a specific case where $b(a)$ is a power-law function,
\ba
2\epsilon\rho=b(a)=c a^{-n} \,.
\ea
where $c$ is a constant. We assume $n>3$ in order to have
the growth on superhorizon scales.

From eq.~(\ref{Ya}) we have 
\begin{align}
a\propto Y^{1/(6-2n)}\,.
\label{aY}
\end{align}
Now eq.~(\ref{dP}) gives 
\begin{align}
P&=\int^a\left(b(a)\frac{da}{a}+\frac{db(a)}{3}\right)
\nonumber\\
&=-\frac{c}{n}a^{-n}+\frac{c}{3}a^{-n}+const.
\nonumber\\
&=\frac{n-3}{3n}b(a)+const..
\end{align}
Plugging eq.~(\ref{aY}) into this, we finally obtain
\be
L=P(Y)= Y^{n/(2n-6)}-V_0 \,. \label{GUSR}
\ee
Since this may be regarded as a natural generalization of the USR case,
which corresponds to the case $n=6$,
we call it the generalized USR (GUSR) model. Lagrangians involving $Y^{\alpha}$ terms have already been studied in \cite{Chen:2013aj,Chen:2013eea,Ginflation}, but those models are either not exactly globally adiabatic because of the presence of a not constant potential or they satisfy the relation $\epsilon\propto a^{-n}$ only approximately and during a limited time range, while for GUSR $\epsilon\propto a^{-n}$ is an exact relation and is valid at any time. As the Lagrangian is of the type described in eqs.~(\ref{condition}) and (\ref{Lag}) (remember that after a field transformation $Y$ can be made equal to $X$), we understand that this scalar field model is indeed equivalent to a barotropic fluid. Hence we have $c_w^2=c_s^2$, and therefore $\delta P_{ nad}=0$. Indeed the second condition for super-horizon growth of $\R_c$ given in eq.~(\ref{GA}) is satisfied.
More precisely, we note that for the GUSR model, the sound velocity is 
exactly constant,
\be
c_w^2=c_s^2=\frac{n-3}{3} \label{cs2new} \,.
\ee

The power spectrum of the comoving curvature perturbation can be
explicitly computed for this model. One finds \cite{NEW}
that the spectral index is a function of $n$: $n_s-1= 6-n$, in agreement with the scale invariant spectrum of the original ultra slow-roll inflation in which one has $n=6$. Hence, the model can be 
constrained by the observational value. Note as well, from eq.~(\ref{cs2new}), that to have a slightly red-tilted spectrum, we need a slightly superluminal speed of sound. 

\subsection{Ex 2: Lambert Inflation}
As another example, let us consider the case when $\epsilon$ 
is a power-law function,
\begin{align}
\epsilon(a)=\epsilon_0{a}^{-n}\,.
\label{powereps}
\end{align}
As before, we assume $n>3$.
In this case, since $d\log\rho/d\log a=-2\epsilon\propto a^{-n}$,
we find
\be
\rho(a)
=\rho_0\exp\left[\frac{2\epsilon}{n}\right]
\label{rhoa} \,.
\ee
It is clear that $\rho$ approaches a constant $\rho_0$
 asymptotically at $a\to\infty$.

Inserting eq.~(\ref{powereps}) and eq.~(\ref{rhoa}) into eq.~(\ref{cs2}),
the sound velocity is given by
\begin{align}
c_w^2=c_s^2=-1-\frac{1}{3}\left(\frac{d\log\epsilon}{d\log a}
+\frac{d\log\rho}{d\log a}\right)
=\frac{n-3+2\epsilon}{3}\,.
\end{align}
Thus $c_s^2$ is time dependent, but it rapidly approaches a constant
as $\epsilon$ decays out.
Also from eq.~(\ref{powereps}) and eq.~(\ref{rhoa}), we find
\begin{align}
b(a)=2\epsilon\rho=2\epsilon\rho_0\exp\left[\frac{2\epsilon}{n}\right]\,.
\label{bform}
\end{align}
Thus we have
\begin{align}
Y\propto a^6 b^2\propto a^{6-2n}\exp\left[\frac{4\epsilon}{n}\right]
\propto \epsilon^{(2n-6)/n}\exp\left[\frac{4\epsilon}{n}\right] \,,
\end{align}
which implies
\begin{align}
Y^{n/(2n-6)}\propto\frac{4\epsilon}{2n-6}
\exp\left[\frac{4\epsilon}{2n-6}\right]\,.
\label{Yeps}
\end{align}

To find the Lagrangian, we manipulate eq.~(\ref{dP}) as
\begin{align}
dP&=b\frac{da}{a}+\frac{db}{3}
=-\frac{b}{n}\frac{d\epsilon}{\epsilon}+\frac{db}{3}
\nonumber\\
&=-\frac{2}{n}\rho_0e^{2\epsilon/n}d\epsilon+\frac{db}{3}\,.
\end{align}
Therefore, integrating this we obtain
\begin{align}
P=\rho_0e^{2\epsilon/n}\left(-1+\frac{2\epsilon}{3}\right)+const..
\label{Pepsilon}
\end{align}
One can invert eq.~(\ref{Yeps}) to find $\epsilon$ as a function
of $Y$, and then insert it into the above to obtain the Lagrangian.

Specifically, we introduce the Lambert function $W(x)$ defined by
the inverse function of $X(z)=ze^z$,
\begin{align}
z=X^{-1}(ze^z)\equiv W(ze^z)\,.
\end{align}
Setting
\begin{align}
{Y}^{n/(2n-6)}=ze^z\,;
\quad z=\frac{4\epsilon}{2n-6}\,,
\end{align}
we have
\begin{align}
\frac{4\epsilon}{2n-6}=W(y)\,;
\quad y\equiv Y^{n/(2n-6)}\,.
\label{yexpY}
\end{align}
Inserting this into eq.~(\ref{Pepsilon}), we finally obtain
\begin{align}
L&=P(Y)
\nonumber\\
&=\rho_0
\left(\frac{n-3}{3}W(y)-1\right)\exp\left[\frac{n-3}{n}W(y)\right]
-V_0\,,
\label{Laminf}
\end{align}
where $y=y(Y)$ is given in eq.~(\ref{yexpY}).

Note that this model has  been derived without making any approximation, 
and it gives exactly $\epsilon \propto a^{-n}$. However, as we mentioned
before, in the late time limit, there is no difference between
$\epsilon \propto a^{-n}$ and $\rho\epsilon \propto a^{-n}$.
Thus the two models discussed above are essentially the same at late times.
This can be easily checked by expanding $W(y)$ around $y=0$,
\begin{align}
W(y)=y-y^2+\cdots.
\end{align}
At leading order in $y=Y^{n/(2n-6)}$, this gives
\begin{align}
P(Y)=\frac{n-3}{3}\rho_0Y^{n/(2n-6)}-\rho_0-V_0\,.
\end{align}
By absorbing the constant coefficient into $g(\phi)$ 
in the definition of $Y$, $Y=g(\phi)X$, and absorbing $\rho_0$ 
into the constant $V_0$, eq.~(\ref{Laminf}) reduces to
\begin{align}
P=Y^{n/(2n-6)}-V_0\,,
\end{align}
which indeed coincides with the GUSR model, see eq.~(\ref{GUSR}).

Higher order terms in the expansion give an infinite class of
 models of the type
\ba
u(Y)&=&\sum_i \beta_i Y^{n_i} \,, \label{fY3}
\ea
where $\beta_i$ are appropriate coefficients.

Finally, note that in USR and as well in the two examples considered here, the shift symmetry in the potential ($V(\phi)=V_0$) is a direct consequence of the demand $c_w^2=c_s^2$, which in turn follows from the global a\-di\-a\-ba\-ti\-ci\-ty of the model. That is in line with the general statement \cite{Akhoury:2008nn,Sawicki:2012re} that for a $k$-essence theory to describe a fluid, one needs a shift symmetry (i.e., there is no physical clock, the model is of the non-attractor type).

\section{Conclusions}
By introducing the notion of global adiabaticity, namely,
$c_w^2=c_s^2$ and $\delta P_{nad}=\delta P-c_w^2\delta\rho=0$,
where $c_w^2=\dot P/\dot\rho$ and $c_s$ is the propagation (phase) speed
of the perturbation, we have determined the general conditions for 
the non-conservation 
of the curvature perturbations on comoving slices $\R_c$ on super-horizon scales.
We have found that globally adiabatic K-essence models can exhibit 
this behavior.

We have then developed a method to construct the Lagrangian of 
a K-essence globally adiabatic (GA) model by specifying
the behavior of background quantities such as 
$\epsilon\rho$ where $\epsilon$ is the slow-roll parameter,
using the equivalence between barotropic fluids and GA K-essence models. 
We have applied the method to find the equations of state of the fluids
and derive the Lagrangian of the equivalent single scalar field models. 
Interestingly, we have found that 
the requirement to avoid the gradient instability, ie, $c_s^2>0$ 
is almost identical to the condition for the non-conservation on
superhorizon scales.

The advantage of our approach is that we have not solved any perturbation 
equation explicitly, since we have proceeded in the opposite way solving 
the inversion problem consisting of requiring certain properties 
to the behavior of the perturbation equation we are interested in. 
In other words, instead of starting from a Lagrangian and then solve the 
perturbations equations we have determined the equation of state or 
equivalently the Lagrangian which admits a solution of the perturbation 
equation with the particular behavior we are interested in.

We have shown that the main difference between attractor models and GA
models is that the latter are adiabatic on all scales, 
while attractor models are approximately adiabatic in the sense
of $\delta P_{nad}=0$ only on super-horizon scales and $c_w^2\neq c_s^2$.

The detailed study of the new models found in this paper will be done in a 
separate upcoming work \cite{NEW} but we can already predict that they can
 be compatible with observational constraints on the spectral index thanks 
to the extra parameter $n$ which is not present in USR. Furthermore they can violate the 
Maldacena's consistency condition and consequently produce large local 
shape non-Gaussianity.

In the future it will be interesting to apply the inversion method we
 have developed to other problems related to primordial curvature
 perturbations, or to develop a similar method for the adiabatic sound 
speed as function of the scale factor.
 
\acknowledgments

The work of MS was supported by MEXT KAKENHI No.~15H05888. SM is funded by the Fondecyt 2015 Postdoctoral Grant 3150126.
This work was supported by %the European Union (European Social Fund, ESF) and Greek national funds under the ?ARISTEIA II? Action.
the Dedicacion exclusica and Sostenibilidad programs at UDEA, the UDEA CODI project IN10219CE and 2015-4044, and Colciencias mobility project COSOMOLOGY AFTER BICEP.

% This work was supported by the European Union (European Social Fund, ESF) and Greek national funds under the ?ARISTEIA II? Action.
% %, the Dedicacion exclusica and Sostenibilidad programs at UDEA, the UDEA CODI
% %project IN10219CE.
% The work of S.G. was supported by the Theoretical Astroparticle Physics research Grant No. 2012CPPYP7 under the Program PRIN 2012 funded by the Ministero dell'Istruzione, Universit\`a e della Ricerca (MIUR).

%\bibliography{Bibliography}
%\bibliographystyle{h-physrev4}

\end{document}